\documentstyle[12pt]{article}
\include{epsf}
\def\lo{\langle 0 |}
\def\ro{ | 0 \rangle }

\def\gmf{\gamma _{5}}

\def\la{\langle }
\def\ra{ \rangle }

\newcommand{\beq}{\begin{equation}}
\newcommand{\eeq}{\end{equation}}
\newcommand{\bea}{\begin{eqnarray}}
\newcommand{\eea}{\end{eqnarray}}

\begin{document}
                                        \begin{titlepage}
\begin{flushright}
hep-th/9802095
\end{flushright}
\vskip1.8cm
\begin{center}
{\LARGE
 ``Integrating in" and Effective Lagrangian  \\       
\vskip0.7cm
for Non-Supersymmetric Yang-Mills Theory    \\
%\vskip1.0cm
 %and 4D Gluodynamics    
}         
\vskip1.5cm
 {\Large Igor~Halperin} 
and 
{\Large Ariel~Zhitnitsky}
\vskip0.5cm
        Physics and Astronomy Department \\
        University of British Columbia \\
 6224 Agricultural Road, Vancouver, BC V6T 1Z1, Canada \\ 
        {\small e-mail: 
higor@physics.ubc.ca \\
arz@physics.ubc.ca }
\vskip1.0cm
{\Large Abstract:\\}
\end{center}
\parbox[t]{\textwidth}{ 
Recently a non-supersymmetric analog of Veneziano-Yankielowicz
(VY) effective Lagrangian has been proposed and applied for 
the analysis of the $ \theta $ dependence in pure Yang-Mills theory.
This effective Lagrangian is similar in many respects to the 
VY construction and, in particular, exhibits a kind of 
low energy holomorphy which is absent in the full YM  theory.
Here we incorporate a heavy fermion into this effective 
theory by using the ``integrating in" technique.
We find that, in terms of this extended theory, holomorphy 
of the effective Lagrangian for pure YM theory naturally
implies a holomorphic dependence
on the heavy fermion mass.
 It is shown  that this analysis fixes, under 
certain assumptions, a dimensionless parameter
which enters the effective Lagrangian and determines the number 
of nondegenerate vacuum
sectors in pure YM theory. We also compare our results 
for the vacuum structure and $ \theta $ dependence to
those obtained recently by Witten on the basis of AdS/CFT 
correspondence.
}

\vspace{1.0cm}

                                                \end{titlepage}

\section{Introduction}

The remarkable progress made over the past years in understanding
the nonperturbative properties of 
supersymmetric (SUSY) N=1 and N=2 theories 
owes a lot to holomorphy. Supersymmetry 
requires a superpotential $ W_{eff}(X_{i},g_{i}) $
of the effective low energy theory to be a holomorphic
function of the light fields $ X_i $ and the coupling constants
$ g_i $, 
and thus powerfully constraints the 
large distance dynamics of SUSY theories \cite{Seib} (see e.g. 
\cite{SUSY} for a review). The famous Veneziano-Yankielowicz (VY)
effective Lagrangian \cite{VY}, obtained from the anomalous 
Ward identities of SUSY YM theory, is holomorphic in its fields
and parameters\footnote{It is well known that the VY effective 
Lagrangian is not a genuine Wilsonian effective Lagrangian
for light degrees of freedom,
but rather has a different meaning, see Sect.2.}. 
Moreover, the effective VY potentials for theories
with different numbers of matter fields (e.g. ones with $ N_f $ and 
$ N_f -1 $ fermion flavors) are related to each other 
by holomorphic decoupling relations \cite{VY}. More recently, 
an operation inverse to integrating out of a heavy fermion
was suggested under the name of ``integrating in" procedure
\cite{Int}. The integrating in technique provides a powerful
method to obtain relations between parameters and functional
forms of superpotentials for theories with different matter
contents. In particular, it was shown in \cite{Int} that 
Affleck-Dine-Seiberg  superpotential \cite{ADS} can be 
obtained from VY Lagrangian for SUSY gluodynamics by 
the integrating in/out procedure. 
The integration in technique was used to study the 
phase structure of SUSY theories \cite{IS}, and to calculate
numerical parameters in different models \cite{FP}.
     
Recently a non-supersymmetric analog of the VY effective Lagrangian
has been proposed \cite{1}
using an infinite series of anomalous Ward 
identities for YM theory. This construction has some 
striking similarities to its supersymmetric counterpart.
Specifically, we have found that this effective Lagrangian
(more precisely, effective potential) possesses both a ``dynamical"
part, which is similar to the original VY form \cite{VY}, and 
a ``topological" part, which is analogous to an improvement 
of the VY Lagrangian, suggested recently \cite{KS}. The picture
of the physical $ \theta $ dependence in pure YM theory, following 
from the analysis of this effective Lagrangian, is rather similar
to the one found for SUSY YM theory \cite{SV,KS}: the correct 
$ 2 \pi $ periodicity in $ \theta $ is recovered when a set of 
disconnected vacua is taken into account. The difference from the 
supersymmetric case is in the absence of degeneracy between the 
vacua. As a result, all vacua but one with a lowest (for a fixed 
$ \theta $) energy contribute zero to the partition function
in the thermodynamic limit \cite{1}. The number of
vacua can be found from explicit dynamical calculations only.
For the SUSY case, such calculations can be done \cite{SV},
while no similar technique is known in the non-supersymmetric 
case.
 The information on the number of vacua
was coded in the effective Lagrangian
of Ref.\cite{1} in a dimensionless number $ \xi \sim N_{c}^{-1} $ 
(see Eq.(\ref{3}) below). (On general grounds, the number of 
vacua in YM theory should be proportional to the number of colors 
$ N_c $.)

Another remarkable property of the effective potential of Ref.\cite{1}
is its holomorphic structure. At first sight, a claim of 
holomorphy of an effective Lagrangian for YM theory may sound 
suspicious, 
as the full theory is not seen to possess any holomorphy, in contrast
to supersymmetric models. As will be discussed in detail below, there
is no contradiction here: holomorphy is the property of 
a nonperturbative effective 
potential (defined within a 
particular regularization scheme)
which is fixed in this approach by the anomalous Ward 
identities and some additional arguments, and describes the large
distance physics only. We do not expect that kinetic terms and/or
perturbative contributions would be holomorphic
as well, but they are irrelevant anyway for our purposes.

In this paper we analyse holomorphy of the effective Lagrangian
of Ref.\cite{1} using the integration in technique \cite{Int}.
By the integrating in procedure we obtain an effective 
Lagrangian for the YM field interacting with a heavy fermion.
It will be shown that holomorphy of the effective Lagrangian
for pure YM theory corresponds to a holomorphic dependence on a 
fermion mass in an extended theory including the heavy fermion.
When gluodynamics is defined as a low energy limit of the latter
theory, the integrating in procedure fixes 
(under certain plausible assumptions specified below) the 
aforementioned
parameter $ \xi $, and thus allows one to find the number of different
nondegenerate $ \theta $ vacuum sectors \cite{1} for pure YM theory.
As will be shown below, this approach results in the same value 
$ \xi = 4/(3b) $ which was obtained in \cite{2} with a different 
method which implies, however, the same
regularization prescription.

Our presentation is organized as follows. In Sect.2 we recall
the construction \cite{1} of the effective Lagrangian for 
pure gluodynamics.
In Sect.3 a new effective 
Lagrangian for the theory including a heavy fermion is obtained 
by the integration in technique, assuming the standard form of the 
fermion mass term and preservation of holomorphy under the procedure 
of integrating in/out.  
We show that the parameter $ \xi $ can be fixed by comparing the 
holomorphic properties of two effective Lagrangians.
The matching of two effective theories is further discussed in Sect.4
from the point of view of their global, ``topological", properties.
We also study in this section a connection
of the present approach with the 
previous analysis of Ref.\cite{2}.  In Sect.5 we 
discuss our results and compare them with different approaches 
to the problems of interest. In particular, we argue that
our picture of the vacuum structure and $ \theta $ dependence 
is in qualitative agreement with that recently obtained by Witten
\cite{Wittheta} in the limit $ N_c \rightarrow \infty $
within a different approach. The final Sect.6 contains our 
conclusions.   

\section{Effective Lagrangian for gluodynamics}

The purpose of this section is to describe an effective Lagrangian
for YM theory, which was constructed in our previous paper \cite{1}.
Before proceeding with the presentation, we would like to pause
for a comment on the meaning of this effective Lagrangian. As 
there exist no Goldstone bosons in pure YM theory, no Wilsonian
effective Lagrangian, which would correspond to integrating out 
heavy modes, can be constructed for gluodynamics. Instead, one 
speaks in this case of an effective Lagrangian as a generating 
functional for vertex functions of the composite fields $ G^2 $ and 
$ G \tilde{G} $. Moreover, only the potential part of this 
Lagrangian can be found as it corresponds to zero momentum n-point
functions of $ G^2 , G \tilde{G} $, fixed by arguments 
appealing to renormalizability of the theory. 
The kinetic part is not fixed 
in this way. Thus, such an effective Lagrangian is not very useful
for calculating the S-matrix, but is perfectly suitable for addressing 
the vacuum properties\footnote{  
For SUSY theories, 
Veneziano-Yankielowicz effective Lagrangian \cite{VY} has the 
same meaning, see \cite{KS}.}. Specifically, space-time
independent fields are amenable to a study within this framework.

Analogously to the case of SUSY theories, an effective Lagrangian
for pure YM theory is constructed using an infinite 
series of anomalous Ward identities which serve as matching conditions 
ensuring consistency
of the large distance properties of the theory with its small 
distance behavior fixed by renormalizability and asymptotic 
freedom. The complete set 
of two-point correlation functions is 
\beq
\label{1}
\lim_{q \rightarrow 0 } \; i 
\int dx \, e^{iqx} \lo T 
\{ \frac{\beta(\alpha_s)}{4 \alpha_s} G^2 (x)  \,  
 \frac{\beta(\alpha_s)}{4 \alpha_s} G^2 (0) \} \ro  = 
-4  \la \frac{
\beta(\alpha_s)}{ 4 \alpha_s} G^2 \ra \; ,  
\eeq
\beq
\label{2}
\lim_{q \rightarrow 0 } \; i \int dx \, e^{iqx} \lo T 
\left\{ \frac{\beta(\alpha_s)}{4 \alpha_s} 
G^2 (x)  \, 
\frac{\alpha_s}{8 \pi} G \tilde{G} (0) \right\}  \ro =  -  
  4 \la \frac{
\alpha_s}{ 8 \pi} G \tilde{G} \ra \; , 
\eeq
\beq
\label{3}
\lim_{ q \rightarrow 0} \; 
i \int dx \, e^{iqx} \lo T \left\{ \frac{\alpha_s}{8 \pi} 
G \tilde{G} (x)  \, 
\frac{\alpha_s}{8 \pi} G \tilde{G} (0) \right\}  \ro =    
 \xi^2  \, \la \frac{
\beta(\alpha_s)}{ 4 \alpha_s} G^2 \ra \; . 
\eeq
Here $ \xi $ is a dimensionless parameter which will be assumed 
to be a rational number. (An irrational value of $ \xi $ would 
presumably produce a non-differentiable $ \theta $ dependence for 
YM theory.)
Multi-point correlation functions of the fields $ G_{\mu \nu}^2 $
and $ G_{\mu \nu} \tilde{G}_{\mu \nu} $ are obtained by 
differentiating Eqs.(\ref{1}-\ref{3}) with respect to $ \theta $ and 
$ 1/g_{0}^2 $, see below.
In what follows we use the one-loop $ \beta $-function
$ \beta(\alpha_s) = - b \alpha_{s}^{2}/(2 \pi) $ where $ 
b = (11/3) N_c $, though most of the discussion below can also 
be formulated with formally keeping the full $ \beta $-function.

A few comments on these Ward identities are in order\footnote{
Although we will occasionally call Eqs. (\ref{1})-(\ref{3}) the 
Ward identities, it should be mentioned that Eq. (\ref{3}) is not 
precisely a Ward identity, but rather is a scheme dependent relation
involving an unspecified at this stage parameter $ \xi $.}.
All correlation functions (\ref{1}),(\ref{2}),(\ref{3}) are defined 
via Wick type of the T-product, i.e. by the differentiation of the 
path integral in respect to corresponding parameters, see below.
Perturbative contributions to the conformal anomaly matrix element
$ \la - b \alpha_s/(8 \pi) G^2 \ra $ in Eqs. (\ref{1}) and (\ref{3})
are subtracted 
to any finite order in $ \alpha_s $ 
by definition. In this case, its dependence on the 
bare coupling constant $ g_0 $ corresponding to the cut-off scale
$ M_R $ is fixed by the dimensional transmutation formula
\beq
\label{4}
\la - \frac{b \alpha_s}{8 \pi} G^2 \ra = c_1  \left[ M_R \exp \left(- 
\frac{8 \pi^2}{b
g_0^2} \right) \right]^4 \equiv c_1 \, \Lambda_{YM}^4 \; .
\eeq
Analogously, $ \la \alpha_s/(8 \pi) G \tilde{G} \ra = c_2 
\Lambda_{YM}^4
$. Here numerical constants $ c_1 , c_2 $ are independent of 
$ g_0 $,
but depend on the vacuum angle $ \theta $ which is allowed to 
be non-zero
in Eqs.(\ref{1}-\ref{3}). 
We note that the constants $ c_1, c_2 $ depend on a particular
regularization scheme used to define the nonperturbative vacuum 
condensates. However, once specified, the VEV (\ref{4}) determines
{\it all} zero momentum correlation functions of $ 
\beta(\alpha_s)/(4 \alpha_s) G^2 $, with perturbative tails subtracted. 
The Ward identities (\ref{1}),(\ref{2})
then follow by the differentiation of the above expressions with 
respect
to $ 1/g_{0}^2 $. They were derived long ago by Novikov, Shifman, 
Vainshtein and Zakharov (NSVZ) \cite{NSVZ}. By derivation, the 
two-point 
functions (\ref{1}-\ref{3}) do not contain perturbative 
contributions\footnote{ The same result (\ref{1}) was 
obtained in \cite{NSVZ} using canonical methods with 
Pauli-Villars regularization. To one loop order in regulator
fields, it was found that perturbative contributions add an identity
to both sides of Eq.(\ref{1}). The absence of perturbative 
contributions 
to the correlation functions (\ref{2}),(\ref{3}) 
is obvious.}. As was discussed 
in detail in Ref.\cite{1}, the zero momentum correlation functions
(\ref{1}-\ref{3}) and their n-point generalizations are generated
by the differentiation of $ \log (Z/Z_{PT}) $ with respect to 
$ 1/g_{0}^2 $ and $ \theta $, where 
\beq
\label{5}
Z (\theta ) =  Z_{PT} \exp \left\{ - i V E_{v}(\theta) \right\}
 = Z_{PT} \, \exp \left\{ - i V  \lo 
- \frac{ b \alpha_s}{ 32 \pi}
G^2 \ro_{\theta} \right\} \; .
\eeq 
Here the perturbatively defined partition function $ Z_{PT} $ does not
depend on $ \theta $ and absorbs perturbative contributions to the 
conformal anomaly. The dimensionless parameter $ \xi $ in Eq.(\ref{3})
is related to the $ \theta $ dependence of the vacuum energy $ E_{v}(
\theta) $ in Eq.(\ref{5}):
\beq
\label{6}
E_{v} (\theta) = E_{v}(0) \, f(\theta) \; \; , \; \; 
f(\theta) = 1 - 2 \xi^2 \, \theta^2 + \cdots \; .
\eeq
Let us introduce complex linear combinations of the composite
fields
\beq
\label{7}
H  = \frac{b \alpha_s}{16 \pi} \left( -
G^2  + i \, \frac{2}{ b \xi} \ 
G \tilde{G} \right) \; , \; \bar{H}   
= \frac{b \alpha_s}{16 \pi} \left( - 
G^2  - i \, \frac{2}{b \xi} 
G \tilde{G} \right) \; .
\eeq  
In terms of these combinations, the Ward identities (\ref{1}-\ref{3}) 
take particularly simple forms (for an arbitrary
value of the vacuum angle $ \theta $):
\bea
\label{8}
\lim_{q 
\rightarrow 0 } \, i \int dx 
e^{iqx} \lo T \{ H(x) \; H(0) \} \ro &=& - 4 \la H \ra \; , 
\nonumber \\
\lim_{ q \rightarrow 
0} \,  i \int dx 
e^{iqx} \lo T \{ \bar{H}(x) \; \bar{H}(0) \} \ro &=& - 4 
\la \bar{H} \ra \; , \\ 
\lim_{ q \rightarrow 
0} \,  i \int dx 
e^{iqx} \lo T \{ \bar{H}(x) \; H(0) \} \ro &=&  0 \; . \nonumber 
\eea
It can be seen that the n-point zero momentum correlation
function of the operator $ H $ equals $ (-4)^{n-1} \la H \ra $.
Multi-point correlation functions of the operator $ \bar{H} $ 
are analogously expressed in terms of its vacuum expectation value 
$ \la \bar{H} \ra $. At the same time,
it is easy to check that the decoupling of the fields $ H $ 
and $ \bar{H} $ 
holds for arbitrary n-point functions 
of $ H $, $ \bar{H} $. This is the origin of holomorphy of an effective
Lagrangian for YM theory\footnote{To avoid possible misunderstanding,
we note that the right hand side of the last equation in (\ref{8})
does contain perturbative contributions proportional to regular 
powers of $ \alpha_s $. However, they are irrelevant for our purposes.
The decoupling of the fields $ H $ and $ \bar{H} $ holds at the level
of nonperturbative $ O(e^{- 1/\alpha_s}) $ effects.  
Holomorphy of an effective potential for YM theory has the same 
status.
Thus, in contrast to the supersymmetric case where holomorphy 
is an exact property of the effective superpotential, in the 
present case it only refers to a ``nonperturbative" effective potential
which does not include perturbative effects to any finite order in 
$ \alpha_s $. We assume that perturbative and nonperturbative 
effects can be separated, at least in principle or/and by some
suitable convention \cite{2,1}.}. 

An effective low energy Lagrangian (more precisely, 
effective potential)
is now constructed 
as the (Legendre transform of) generating functional for zero 
momentum
correlation functions of the marginal operators $  G_{\mu \nu} 
\tilde{G}_{\mu \nu} $ and $ G_{\mu \nu} 
G_{\mu \nu} $, such as Eq.(\ref{8}) and their n-point generalizations.
It is a function of effective zero momentum fields $ h , \bar{h} $
which describe the vacuum expectation values of the fields 
$ H ,\bar{H} $ :
\beq
\label{9}
\int dx \, h = \la \int dx \, H \ra \; \; , \; \; 
\int dx \, \bar{h} =  \la \int dx \, \bar{H} \ra \; .
\eeq
Omitting details of the derivation, which 
can be found in \cite{1}, we give the final answer for the 
improved effective potential $  
F(h,\bar{h}) $:
\bea
\label{10}
e^{- i V F(h,\bar{h}) } &=& \sum_{n = - \infty}^{
 + \infty} \sum_{k=0}^{q-1} \exp \left\{ - \frac{i V}{4}
\left( h \, Log \, \frac{h}{C} + 
\bar{h} \, Log \, \frac{ \bar{h}}{
\bar{C}} \right) \right. \nonumber \\ 
&+& \left. i \pi V \left( k + \frac{q}{p} \,  
\frac{ \theta + 2 \pi n}{ 2 \pi} \right) \frac{h - \bar{h}}{
2 i} \right\} \;  ,
\eea
where    
the constants $ C, \bar{C} $ can be set real and expressed in terms 
of the vacuum energy at $ \theta = 0 $, 
$  C =  \bar{C} = - 2 e E_{v}(0) $, see Eq.(\ref{6}), and $ V $ 
is the 4-volume. The holomorphic structure of the effective 
potential is explicit in Eq.(\ref{10}).
The integer numbers $ p $ and $q $ are relatively prime.
As was shown in Ref.\cite{1}, they are related to the 
parameter $ \xi $ introduced in Eq.(\ref{3}), $ 2 \xi = q/ p $. 
The symbol $ Log $ in 
Eq.(\ref{10}) stands for the principal branch of the logarithm.
The effective potential (\ref{10}) produces an infinite series
of anomalous WI's. By construction, it is a periodic
function of the vacuum angle $ \theta $.  
The effective potential 
(\ref{10})  
is suitable for a study of the YM vacuum as described above.

The double sum over the integers 
$ n , k $ in
Eq.(\ref{3}) appears as a resolution of an ambiguity
of the effective potential as defined from the anomalous WI's.
As was discussed in \cite{1}, this ambiguity is due to the fact
that any particular branch of the multi-valued function
\beq
\label{11}
\log \, z^{p/q} = \frac{p}{q} \, Log \, z + 2 \pi i ( n + k 
\frac{p}{q}) \; \; , \; \;  n = 0, \pm 1,\ldots \; \; ; \; 
\; k = 0, 1, \ldots , q-1 \; 
\eeq
 corresponding to some fixed 
values of $ n , k $, satisfies the anomalous WI's.
However, without the summation over the integers $ n , k $ 
in Eq.(\ref{10}), the effective potential would be   
multi-valued and unbounded from below. 
An analogous problem arises with the 
original VY effective Lagrangian. It was cured 
by Kovner and Shifman in \cite{KS}
by a similar prescription of summation over all branches of the 
multi-valued VY superpotential. Moreover, the whole 
structure of Eq.(\ref{10}) is rather similar to that of the (amended)
VY effective potential. Namely, it contains
both the ``dynamical" and ``topological" parts (the 
first and the second terms in the exponent, respectively).
The ``dynamical" part of the effective potential (\ref{10}) 
is similar to the VY \cite{VY}
potential $
\sim S \log (S/ \Lambda)^{N_c} $ (here 
$ S $ is an anomaly superfield), 
while the ``topological" part is 
akin to the improvement \cite{KS} of the VY effective potential. 
Similarly to the supersymmetric case, the infinite sum over $ n $ 
reflects
the summation over all integer topological charges in the 
original YM theory. The difference of our case from 
that of supersymmetric YM theory is that 
an effective potential of the form $ (1/N) \phi \log (\phi/\Lambda)^N
$ , as in the SUSY case, implies a simpler form of the 
``topological" term $ \sim 2 \pi i n /N \, (\phi - \bar{\phi}) $
with only one ``topological number" $ n $ which specifies the 
particular branch of the multi-valued logarithm. In our case,
we allow for a more 
general situation when the parameter $ N $ is a rational number 
$ N = p/q $. In this case we have two integer 
valued ``topological numbers"
$ n $ and $ k $, specifying the branches of the logarithm and 
rational function, respectively.
Our choice is 
related to the fact that some proposals to fix the values of $ p, q $
suggest that $ q \neq 1 $. One may expect
the integers $ p $ and $ q $ are related to
a discrete symmetry surviving the anomaly,
which may not be directly visible in the original fundamental
Lagrangian. Our purpose in this paper is to try to find these 
numbers by analysing the properties of the effective potential
(\ref{10}).

It should be stressed that the improved effective potential
(\ref{10}) contains more information in comparison to that 
present in the anomalous Ward identities just due to the 
presence of the ``topological" part in Eq.(\ref{10}).
Without this term Eq.(\ref{10}) would merely be a kinematical
reformulation of the content of anomalous Ward identities for 
YM theory. The reason is that the latter refer,  
as usual, to the infinite volume (thermodynamic) limit 
of the theory, where only one state of a lowest energy (for $ 
\theta $ fixed) survives. This state corresponds to one 
particular branch of the multivalued effective potential
in Eq.(\ref{10}). At the same time, the very fact of multi-valuedness
of the effective potential implies that there are other vacua which 
should all be taken into consideration when $ \theta $ is varied.
When summing over the integers $ n \, , \, k $, we keep track of 
all (including excited) vacua of the theory, and 
simultaneously solve the problems
of multi-valuedness and unboundedness from below of the ``one-branch
theory". The most attractive feature of the proposed structure
of the effective potential (\ref{10}) is that the same summation
over $ n , k $ reproduces the topological charge quantization
and $ 2 \pi $ periodicity in $ \theta $ of the original YM theory.

\section{Integrating in the heavy fermion}

In this section the effective Lagrangian (\ref{10}) for pure 
YM theory will be analysed from a different standpoint. Considering 
pure gluodynamics as a low energy limit of a theory describing 
the YM field interacting with a heavy fermion, we now wish 
to construct an effective Lagrangian for the latter theory
starting from the effective Lagrangian (\ref{10}). Our purpose here
is to try to understand, in this framework, holomorphy 
of the effective potential (\ref{10}) and the constraints imposed by it.
As will be 
argued below and in Sect.4, a relation between the 
holomorphic and ``topological"
properties of two Lagrangians is non-trivial, and allows one 
to fix the crucial parameter $ \xi = q/(2p) $ 
entering Eq.(\ref{10}), under two plausible assumptions, see below.
% A simple argument fixing the value of
%$ \xi $ will be presented below in this section.

The task of constructing such an effective Lagrangian for the theory
with a heavy fermion is  
achieved by using the ``integrating in" technique, developed in the 
content of SUSY theories in Ref.\cite{Int} and reviewed by 
Intriligator and Seiberg in \cite{SUSY}.
The integrating in procedure can be viewed as a method of introducing
an auxiliary field into the effective Lagrangian for pure YM theory.
Using the renormalization group properties of the 
YM effective Lagrangian, the latter is extended to include the auxiliary 
field $ U $, which will be later on identified with the chiral
combination $ \bar{\Psi}_L \Psi_R $ of a heavy fermion.

To conform with the notation and terminology of Ref.\cite{Int},
we will call pure YM theory and the theory with a heavy fermion
the d-theory (from ``downstairs")
and the u-theory (from
``upstairs"), respectively. The 
effective potential of the d-theory is then
$ W_d + W_{d}^{+} $ with 
\beq
\label{12}
W_d = 
\frac{1}{4} \,  \frac{q}{p} \, h \log \left( 
\frac{ h}{ C} \right)^{p/q}
= \frac{1}{4} \, \frac{q}{p} \, h \log \frac{ h^{p/q}}{ (c 
\Lambda_{YM} )^{p/q}}   
\eeq
(here $ c $ is a dimensionless numerical coefficient),   
and the summation over all branches 
of the logarithm in the partition function is implied. In this
section Eq.(\ref{12}) will be understood as representing 
a branch (section) of the multi-valued effective potential,
which corresponds to a lowest energy state for small $ \theta \ll 
\pi $. As was shown in \cite{1}, this section corresponds to 
the principal branch of the rational function in the  
logarithm in Eq.(\ref{12}). In
this case the vacuum expectation value $ \la H \ra $ depends on $
\theta $ as follows:
\beq
\label{th}
 \la H \ra_{\theta} = \la H \ra_{0} e^{2 i \xi \theta } 
\eeq  
We now want to relate \cite{SV,Int} 
the dimensional transmutation parameter
$ \Lambda_{YM} $ of pure YM theory to the scale parameter 
$ \Lambda_{QCD} $ of the u-theory including a heavy quark of mass
$ m  \gg \Lambda_{QCD}, \Lambda_{YM}$. 
We assume both parameters to be defined in the $ 
\overline{MS} $ 
scheme, in which no threshold factors arise in 
corresponding matching conditions.
 The matching condition then follows from 
the standard one-loop relations
\bea
\label{lam}
\Lambda_{YM} &=& M_0 \exp \left( - \frac{8 \pi^2}{ b_{YM} 
g^{2}(M_{0})}
\right) \; \; , \; \; b_{YM} \equiv b = \frac{11}{3} \, N_c \; , 
\nonumber \\
\Lambda_{QCD} &=& M_0 \exp \left( - \frac{8 \pi^2}{ b_{QCD} 
g^{2}(M_{0})}
\right) \; \; , \; \; b_{QCD} = 
\frac{11}{3} \, N_c - \frac{2}{3} \; , 
\eea
and 
the requirement
that the coupling constants of the d- and u- theories coincide
at the decoupling scale $ M_{0} = m $. We obtain
\beq
\label{13}
\Lambda_{YM}^4 = \Lambda_{QCD}^{4} \left( \frac{m^2}{ 
\Lambda_{QCD}^{2}}
\right)^{ 4/(3 b) } \; , 
\eeq
As was explained in Refs.\cite{SV,Int,SUSY}, Eq.(\ref{13}) reflects
the fact that,
for fixed $ \Lambda_{QCD} $, the 
scale parameter $ \Lambda_{YM} $ characterizes the low energy theory
surviving below the scale $ m $, and thus depends on $ m $.  
In this sense, the constant in the logarithm 
in Eq.(\ref{12}) also depends on $ m $ 
\beq
\label{14}
 ( c \Lambda_{YM}^{4})^{p/q} = 
( c \Lambda_{QCD}^{4})^{p/q}  
\, \left( \frac{m}{ \Lambda_{QCD} } \right)^{
8 p /(3 b q) } \; .
\eeq
Following Ref.\cite{Int},
we now wish to consider (a particular branch of) the effective 
potential (\ref{12}) as the result of integrating out the 
auxiliary field $ U $ in the new effective potential 
$ W \equiv W_u - m U $ which corresponds to the u-theory:
$ W_{d}(h,m) = W(h , m , \la U \ra ) $, or
\beq
\label{15}
W_d = [ W_u - m U ]_{ \la U \ra }  \; ,
\eeq
where $ \la U \ra $ is a solution of the classical equation of 
motion for the auxiliary field $ U $ :
\beq
\label{16}
\frac{ \partial W_u}{ \partial U } - m = 0 \; .
\eeq
(At this stage, Eq.(\ref{15}) is merely a definition; later on, 
the term $ m U $ will be identified with the fermion mass term
in the effective Lagrangian.)
Let us note that, according to Eq.(\ref{15}), 
$ W_d $ should depend holomorphically 
on $ \la U \ra $. Our assumption is that this is only possible
if an effective potential $ W $ of the u-theory is itself holomorphic 
in the field $ U $. In 
our opinion, this assumption appears to be 
quite reasonable\footnote{We are not aware of any counter-example 
where a holomorphic d-potential would be obtained from a 
non-holomorphic
u-potential by the integrating out procedure.}. 
Furthermore, one can see  
that Eqs. (\ref{15}),(\ref{16}) actually define the 
potential $ W_d $ as the Legendre transform of $ W_u $.
Therefore we can find the unknown function $ W_u $ from the 
known potential $ W_d $ by the inverse Legendre transform: 
\beq
\label{17}
W_u = [ W_d + m U ]_{ \la m \ra } \; , 
\eeq
where $ \la m \ra $ solves the equation
\beq
\label{18}
\frac{ \partial}{ \partial m } \, ( W_d + m U) = 0 \; .
\eeq
Eq.(\ref{18}) can be considered as an equation of motion for 
the auxiliary ``field" $ m $. It is important to note 
that Eqs. (\ref{15} - \ref{18}) 
imply that $ m $ should be treated as a complex parameter to 
preserve the holomorphic structure of Eq.(\ref{12}).
When substituted in Eq.(\ref{17}),
a solution $ \la m \ra $ of Eq.(\ref{18}) defines the potential
$ W_{u}( h , U , \la m \ra ) $. When this function is found, 
the effective potential $ W $ of the u-theory is defined by the 
relation
\beq 
\label{19}
W(h , U , m ) = W_{u}(h , U , \la m \ra) - m U \; , 
\eeq
in accord with Eq.(\ref{15}).

The solution of Eq.(\ref{18}) is easy to find using Eqs. (\ref{12}),
(\ref{14}) :
\beq
\label{20}
\la m \ra = \frac{2}{3b} \, \frac{h}{U} \;. 
\eeq
Thus, Eq.(\ref{17}) yields
\beq
\label{21}
W_u = - \frac{1}{4} \, \frac{q}{p} \, h \log \left[ 
\left( \frac{c \Lambda_{QCD}^{4}}{h} \right)^{p/q} \left(
\frac{2}{3b} \, \frac{h}{ \Lambda_{QCD} U } \right)^{8p/(3bq)} \right]
+ \frac{2}{3b} \, h \; .
\eeq
Finally, Eq.(\ref{19}) results in the effective potential 
of the u-theory
\beq
\label{22}
W =  \frac{1}{4} \, \frac{q}{p} \, h \log \left[ \left( \frac{
h}{ c \Lambda_{QCD}^{4}} \right)^{ ( 1 - 8/3b)p/q }
\left( \frac{3b}{2} \, \frac{ U}{ c \Lambda_{QCD}^3 } 
\right)^{8 p/(3bq)} \right] + \frac{2}{3b} \, h - m U \; .
\eeq
We expect this effective Lagrangian to describe YM theory
coupled to a heavy quark, corresponding to the field $ U $, 
such that integrating out $ U $ brings us back to the 
effective Lagrangian (\ref{12}) for pure gluodynamics.
Indeed the equation of motion for the field
$ U $ stemming from the effective potential (\ref{22}) reads
\beq
\label{23}
m \la U \ra = \frac{2}{3b} \, h \; .
\eeq
Inserting this classical vacuum expectation value (VEV) back to 
Eq.(\ref{22}) (i.e. integrating out the field $ U $), 
we reproduce the effective potential of the d-theory, Eq.(\ref{12}).
Note that as Eq.(\ref{23}) should preserve the $ N_c $ counting
rule, we obtain $ \la h \ra \sim N_{c}^2 $ , $ b \sim N_c $, 
$ \la U \ra \sim N_c $. The $ N_c $ dependence 
of the VEV $ \la U \ra $ is consistent with the identification
$ \la U \ra \sim \la \bar{\Psi}_L \Psi_R \ra $ which will 
be suggested below. 
 
Analogously to the effective potential (\ref{12}) of the 
d-theory, the new potential (\ref{22}) is not a single-valued
function. The single-valuedness should be imposed, as was done
in Eq.(\ref{10}), by the summation over all branches of $ W $ in the
partition function. This procedure
will be considered in detail in the next section, while here 
we would like to identify the field $ U $ of the 
effective theory with a corresponding operator of the fundamental 
theory. As is seen from (\ref{23}), $ U $ has dimension 3, and thus
should 
describe the VEV of an
operator bilinear in the heavy quark fields. Furthermore, as long as
$ m $ is effectively considered as a complex parameter, this 
operator can only be $ \bar{\Psi}_L \Psi_R $ or $ 
\bar{\Psi}_R \Psi_L $, in accord with the structure of the mass 
term in the underlying fundamental theory. (Here the second assumption
of the present approach is implicit: we assume the standard form 
of the fermion mass term in the effective potential (\ref{22}).) 
To find the 
exact correspondence, we note that, when
the VEV of the field $ h $ is chosen to correspond
to a lowest energy state for $ \theta < \pi $, Eq.(\ref{23}) implies
\bea
\label{24}
\la m U + \bar{m} \bar{U} \ra_{\theta} = - \la \frac{ \alpha_s}{12
\pi} G^2 \ra_{0} \cos \left( 2 \xi \theta \right) \; ,
\nonumber \\
\la m U - \bar{m} \bar{U} \ra_{\theta} = - i \la \frac{ \alpha_s}{12
\pi} G^2 \ra_{0} \sin \left( 2 \xi \theta \right) \; ,
 \eea
which can be compared with the relations\footnote{
The equations (\ref{25}) follow from the operator 
product expansions $ \la m \bar{\Psi} \Psi \ra = - \la \alpha_s /(12 
\pi) G^2 \ra + O(1/m^2) $ , $  \la m \bar{\Psi} i \gmf \Psi 
\ra =  \la \alpha_s /(8  
\pi) G \tilde{G} \ra + O(1/m^2) $, and Eq.(\ref{th}).} 
 between the VEV's in the
underlying theory:
\bea
\label{25}
\la m \bar{\Psi}_L \Psi_R  
 + m  \bar{\Psi}_R \Psi_L \ra_{\theta} = - \la \frac{ \alpha_s}{12
\pi} G^2 \ra_{0} \cos \left( 2 \xi \theta \right) 
+ O \left( \frac{1}{m^2} \right) \; ,
\nonumber \\
\la   m \bar{\Psi}_L \Psi_R  
 - m  \bar{\Psi}_R \Psi_L \ra_{\theta} = - i 
\frac{3b}{4} \, \xi \, \la \frac{ \alpha_s}{12
\pi} G^2 \ra_{0} \sin \left( 2 \xi \theta \right) 
+ O \left( \frac{1}{m^2} \right) \; .
\eea
Comparing Eqs. (\ref{24}) and (\ref{25})
and using the relation $ 2 \xi = q/p $ \cite{1}, we conclude that 
\beq
\label{26}
\xi = \frac{4}{3b} \; \; , \; \; \frac{q}{p} = \frac{8}{3b} \; \; 
; \; \; \la U \ra = \la  \bar{\Psi}_L \Psi_R \ra \; .  
\eeq
We thus see that the introduction of the heavy quark into the
effective theory fixes the parameter $ \xi $ which enters
the effective Lagrangian (\ref{12}) for pure YM theory.
The value obtained coincides with the one suggested by us 
previously within a different method \cite{2}, based on a 
different assumption.
(The correspondence between the two approaches to calculate 
the crucial parameter $ \xi $ will be further 
discussed in the next section.)
Moreover, the comparison of Eqs.  (\ref{24}) and (\ref{25})
shows that the chiral field $ U $ corresponds to the 
chiral fermion bilinear of the fundamental Lagrangian:
\beq
\label{27}
m U \Leftrightarrow m  \bar{\Psi}_L \Psi_R \; \; , \; \; 
 \bar{m} \bar{U}  \Leftrightarrow m  \bar{\Psi}_R \Psi_L \; .
\eeq
This correspondence between the operators of the effective and 
underlying theories has the same meaning as Eqs.(\ref{9}), i.e.
the classical field $ U $ describes the VEV of the chiral combination
$ \bar{\Psi}_L \Psi_R $ of the full theory.

It may be instructive to discuss the result (\ref{26}) in a 
slightly different and more intuitive way. Using Eqs.(\ref{25}),
we can write
\beq
\label{hol}
\la m U \ra 
\sim \la  m  \bar{\Psi}_L \Psi_R \ra = - \frac{1}{2} \left(
1 + \frac{ 3b}{4} \xi \right) \la \frac{ \alpha_s}{12 \pi} G^2 
\ra e^{2 i \xi \theta} - 
 \frac{1}{2} \left(
1 - \frac{ 3b}{4} \xi \right) \la \frac{ \alpha_s}{12 \pi} G^2 
\ra e^{- 2 i \xi \theta} \; , 
\eeq
which is a superposition of the holomorphic ($ \sim \exp (2 i \xi 
\theta ) $) and anti-holomorphic ($ \sim \exp (- 2 i \xi 
\theta ) $) functions. At the same time, the equation of motion
(\ref{23}) requires $ \la U \ra $ to be 
a holomorphic function, as $ \la h \ra $
is such a function. This is only possible, as is seen from 
Eq.(\ref{hol}), when $ 1 - (3b/4) \xi = 0 $ which is equivalent to
(\ref{26}). In other words, the requirement of holomorphy for the 
u-theory singles out parameters of the d-theory.

We would like to pause here to discuss the following issue.
The  main result of our calculations, eq. (\ref{26}) is 
the direct consequence of two fundamental principles. The
first principle is  a modified definition   
 of  the path integral  when the prescription
of  summation over all topological sectors is introduced 
both at the fundamental and effective Lagrangian level, see 
Eq.(\ref{10}).
As was discussed earlier, this definition does not change
any local properties of the theory  (in particular, it does not 
alter the 
WI's) but drastically 
changes the global properties of the theory. The prescription
of summation over the topological classes will be further 
discussed in Sects. 4 and 5. The second 
fundamental principle we adopted from the SUSY theories
is the property of holomorphy in the form of a natural requirement
to have a holomorphic u-theory if a d-theory satisfies
this property (and vice versa).
However, in the SUSY case these two principles are argued to
lead to a suspicious new chiral invariant
vacuum, whose analog certainly can not exist in QCD as we know
from experiment. The question is:
How is it possible that a similar  derivation and prescription
adopted for QCD does not lead to
the new chiral invariant vacuum advocated in the SUSY case\footnote{
We thank the Referee for bringing our attention
to this potential problem.}? 

Our view of this problem is that we believe that in both cases 
(supersymmetric and non-supersymmetric) this suspicious
state does not appear as a well-defined vacuum state;
however in the SUSY case the situation
might be less clear than in the ordinary QCD, see below. The argument
is the following.
Formally, such a state seemingly does appear
as a solution of the equation of motion stemming
from the effective Lagrangian. However, it does 
not appear to be a state stable against quantum
fluctuations. Indeed, 
the modulus of the ``order parameter" $ \phi = \la \lambda \lambda
\ra  $ vanishes in this 
candidate ``vacuum". However, the phase of $ \phi $ is ill-defined 
at this point, and the matrix of second derivatives 
describing quantum fluctuations is also 
ill-defined.
An additional source of ambiguity of the matrix of second derivatives
is a freedom of redefinitions  
 $|\phi| \rightarrow |\phi|^n$ 
or even $| \phi| \rightarrow\log( |\phi| )$ in the effective Lagrangian,
which are able to change the sign of $ W_{|\phi|}'' $ for the
candidate ``vacuum" $ | \phi | = 0 $. This freedom of redefinitions
of $ | \phi | $ is due to our lack of knowledge of the 
kinetic term in the effective SUSY Lagrangian, which otherwise would 
fix the correct definition of canonical field.
 Such a behavior of the 
effective potential is an indication that this state is not 
a genuine 
 vacuum state stable against quantum fluctuations. In fact, the 
situation for the SUSY theories is a little bit more controversial
because one could argue that in the SUSY case 
the vacuum energy is zero and thus all quantum fluctuations
should 
cancel out, no matter
what the eigenvalues  for specific fluctuations are.
Nevertheless, we believe that an accurate analysis
of quantum fluctuations in the SUSY case in the background 
of the  chiral invariant
vacuum state within some suitable regularization
would demonstrate that
this state  should be dismissed as an inappropriate candidate
for the vacuum state.
 
To end up this section, we would like to discuss an apparent 
problem related to the effective potential (\ref{22}). 
Proceeding by analogy with SUSY theories, we might naively expect 
that the effective potential, obtained for large $ m \rightarrow 
\infty $, could be continued to small masses $ m \leq \Lambda $,
where the heavy ``glueball" fields $ h , \bar{h} $ could be 
integrated out. (This is e.g. how the Affleck-Dine-Seiberg
superpotential \cite{ADS} was obtained in Ref.\cite{Int}
from the VY effective potential for SUSY gluodynamics.) However,
such a procedure gives correct results for 
supersymmetric theories just due to specific Ward identities 
which allow one to prove 
that the dependence of the gluino condensate on the 
chiral superfield mass
$ m $, viz. $ \la \lambda \lambda \ra \sim \sqrt m $
for SU(2) gauge group, obtained 
for SQCD for small $ m \ll \Lambda $, is actually exact and 
valid also for large $ m \rightarrow \infty $ \cite{SV}. 
As no such relation holds in the non-supersymmetric 
case, we actually have no reason to continue the above formulas 
to the region of small $ m  \simeq \Lambda $. If we still do so,
one can easily see that we do not reproduce in this way the anomalous 
term in the effective chiral Lagrangian of Ref.\cite{Wit2}. This 
seeming problem with Eq.(\ref{22})
is resolved by the fact that the ``QCD limit" $ m \rightarrow 0 $
and the ``YM limit" $ m \rightarrow \infty $ are actually separated
by a kind of phase transition which changes the number of vacua in 
the theory. As the number 
of vacua is determined by the $ \beta $-function
(see below in Sect. 4), it should not change when a very heavy fermion
is added to pure YM theory, and the effective theory below the 
decoupling scale $ M_{0} \simeq m $ is considered. 
(This requirement will be given a formal content in the 
next section.) On the 
contrary, when the light quarks are present, the $ \beta $-function
changes, and so does the number of vacua. Thus,
analytic continuation of the above formulas 
to the small $ m \ll \Lambda $ region leads to a theory other
than the one described by the effective chiral Lagrangian of 
Ref.\cite{Wit2}. As is shown in \cite{QCD}, the latter is 
correctly reproduced by a different procedure.
Namely, one should start directly from an effective Lagrangian for QCD
with light quarks, which realizes at the tree level 
the anomalous conformal and chiral symmetries of QCD. The ``glueball"
part of this effective Lagrangian for QCD \cite{QCD} 
is similar to Eq.(\ref{10}).

\section{Global quantization, holomorphy, and Ward identities
in the effective Lagrangian approach}

In this section we would like to discuss a few related topics.
First, we analyse the construction of the effective potential 
(\ref{22}) of the u-theory while keeping track of multi-valuedness
of effective potentials for both d- and u-theories. 
As we will see shortly, this analysis is consistent with the 
result (\ref{26}). Another
purpose of this section is to discuss holomorphy of the 
effective theory in the fermion mass $ m $ from the 
point of view of the underlying QCD Lagrangian. This will 
allow us to establish the correspondence with an alternate 
method which was suggested by us earlier in Ref.\cite{2}
to find the parameter $ \xi $, where the same value $ \xi= 
4/(3b) $ was obtained. It will be shown how the results
of Ref.\cite{2} follow from the effective Lagrangian approach.

The analysis of the previous section was based on dealing with 
a fixed branch of the effective potential. 
As
was discussed for SUSY theories in \cite{KS}, and for 
pure YM theory in \cite{1}, the
multi-valuedness of the effective potential 
necessitates the summation 
over all branches of a multi-valued action in the path integral
(this is the prescription for constructing an improved 
effective potential).
This procedure enforces some global quantization rules in 
the effective theory \cite{KS,1} which reflect quantization 
of the topological charge in the fundamental theory. Technically,
the global quantization in the d-theory arises due to the 
Poisson formula
\beq
\label{28}
\sum_{n} \exp \left( 2 \pi i \, n \, V \, \frac{q}{p} \,  
\frac{h- \bar{h}}{4i} \right)
= \sum_{m} \delta \left(  V \, \frac{q}{p} \, \frac{h - 
\bar{h}}{4i} - m \right) \; .
\eeq
Let us now consider the problem of matching the d- 
and u- theories from the point of view of the integrating out 
procedure in the u-theory
with account for the fact of multi-valuedness
of the resulting potential. Given any branch of a multi-valued
action, which is compatible with the Ward identities
of the d-theory and the renormalization group, one achieves
the single-valuedness of the partition function for the u-theory
by the summation over all branches of the action in the path 
integral.
In other words, we impose a universal rule for defining 
the partition function (improved effective potential) 
for both the d- and u-theories.
Furthermore, it is natural to require that, after integrating
out the field $ U $ in such a partition function for the u-theory,
we should come back to the correct partition function of the 
d-theory. This requirement has a non-trivial content, since 
generically the logarithms
in Eqs. (\ref{12}) and (\ref{22}) imply different global 
structures in the complex $ h $ plane, i.e. different 
global quantization rules for the d- and u-theories. 
In this way we extend the integrating in/out procedure
to match not only the perturbative scale redefinition in the 
low energy d-theory, but also the global, topological, properties
of the latter. The meaning of this requirement is that 
{\bf turning on the heavy quark should not change the number 
of $ \theta $
vacuum sectors}. 
As we will see shortly, this ``topological"
matching of the two theories agrees with the result (\ref{26}) for 
the parameter $ \xi $.

Let us define the improved effective potential $ \tilde{F} $ for the 
u-theory in the same way as was done in Eq.(\ref{10}). Using
Eq.(\ref{22}), we obtain
\bea
\label{29}
e^{- i V \tilde{F} }  =  \sum_{n,m = - \infty}^{
 + \infty} \sum_{k,l} \exp \left\{ - iV 
 \left[  \frac{1}{4} (1 - \frac{8}{3b} ) h \, Log \, 
 \frac{h}{c \Lambda_{QCD}^4} 
 + \frac{2}{3b} \, h \, Log  \frac{3b}{2} \,
\frac{U}{c \Lambda_{QCD}^3 }   + 
(h.c.) \right. \right. \nonumber \\
\left. \left.  +   \frac{2}{3b} \, (h + \bar{h}) - 
m U - \bar{m}\bar{U}
%\right. \right. \\   
%- \left. \left. 
- \pi \, 
\frac{q}{p} \left( k \, \frac{p}{q} \, ( 1 - 
\frac{8}{3b} ) + l \, \frac{p}{q} \, \frac{8}{3b} + 
+ n + m + \frac{\theta}{2 \pi} \right)  \,  
 \frac{h - \bar{h}}{2 i} \right] \right\} 
\eea
where the sums over $ k, l $ are finite.
% and $ \theta_{nm} \equiv n + m + \theta/(2 \pi) $. 
It is clear from 
Eq.(\ref{29}) that the minimization condition $ \tilde{F}_{U}' 
= 0 $
leads to the same Eq.(\ref{23}). The reason for that is the 
absence of global quantization rules for the field $ U $ which
is thus unconstrained. 
It can be readily seen that 
substituting this solution back to Eq.(\ref{29}), we reproduce
exactly the ``dynamical" term in Eq.(\ref{10}). However, as 
we stated above, the ``topological" term for the u-theory should 
also match the one of the d-theory. Let us now compare these 
two expressions. For the u-theory, the ``topological" term
is 
\beq
\label{30}
i V \pi \, \frac{q}{p} \left( k \, \frac{p}{q} \, ( 1 - 
\frac{8}{3b} ) + l \, \frac{p}{q} \, \frac{8}{3b} + 
n + m + \frac{ \theta}{2 \pi} \right)  \,  
 \frac{h - \bar{h}}{2 i} \; , 
\eeq
where the integers $ n , m $ are ranged from $ (-\infty) $ to 
$ (+ \infty) $, while the integers $ k , l $ reside on finite 
intervals. On the other hand, the ``topological" term of the 
d-theory is 
\beq
\label{31}
i V \pi \, \frac{q}{p}  \left( k' \, \frac{p}{q}  
+ n'  + \frac{ \theta}{ 2 \pi} \right)  \,  
 \frac{h - \bar{h}}{2 i} \; \; , \; \; n'= 0,\pm 1, \ldots \; \; , 
\; \; k'= 0,1, \ldots , q-1 \; .
\eeq
One can see that expression (\ref{30}) has 
the same form as (\ref{31}) only if 
\beq
\label{32}
% \frac{p}{q} = \frac{3b}{8} \; \; , \; \; 
\frac{p}{q} \, \frac{8}{3b} 
= r = integer \; , 
\eeq
for which expression (\ref{30}) becomes
\beq
\label{33}
i V \pi \, \frac{8}{3br} \left( k \, \frac{3b}{8}   
+ n + m + r( l - k )  + \frac{ \theta}{2 \pi} \right)  \,  
 \frac{h - \bar{h}}{2 i} \; .
\eeq
Therefore, unless the constraint 
(\ref{32}) is satisfied, integrating out
the field $ U $ in the u-theory does not lead back to the
correct d-theory, but instead yields a theory with a different 
global quantization rule for the field $ h $
(i.e. different quantum mechanically). It is remarkable that
these ``topological" arguments are consistent with the 
value of $ \xi $ which was arrived upon in Eq.(\ref{26}) 
(still, they are less restrictive, as it usually happens 
for considerations based on topological arguments, 
than Eq.(\ref{26}) above).
On the other hand,
in contrast to the argument of the previous section, no reference
to the underlying theory was made above; only the internal
consistency of the method just developed was used. The agreement 
between these two different lines of reasoning thus supports both 
the reliability of the calculated value of $ \xi $, and 
self-consistency of the effective Lagrangian constructed.

The next topic we would like to discuss is holomorphy of 
the effective Lagrangians. We have seen in the previous section
that the holomorphic structure of the effective potential for 
the d-theory implies holomorphy in the fermion mass $ m $ 
of the effective potential for the u-theory. This means that 
$ m $ should be effectively treated as a complex parameter, 
with the correspondence rule (\ref{27}) between the operators 
of the effective and fundamental theories. Here we would like 
to discuss what such kind of holomorphy implies in terms of 
the fundamental QCD Lagrangian.

Let us first note that in our case holomorphy refers to 
a physical heavy fermion with mass $ m \rightarrow \infty $.
In the fundamental theory, the introduction of a heavy
fermion of mass $ m $ requires a regularization on yet much 
higher ultra-violet scale $ M_{R} $, which can be 
thought of as the mass of a Pauli-Villars regulator.
 In the infinite mass limit $ m , M_{R} \rightarrow \infty $
the properties of the theory in respect to the physical
and regulator fermions 
are identical (up to some sign 
differences) by definition. Thus, we may
expect the fundamental theory to be 
holomorphic, in a sense, in 
the Pauli-Villars regulator mass $ M_{R} $. The analysis of 
this paper shows what kind of holomorphy we 
may expect: it should be holomorphy of nonperturbative 
vacuum condensates or zero momentum correlation functions
with perturbative tails subtracted, because these are the objects
generated by the effective Lagrangian\footnote{ As 
was mentioned in the 
Introduction, we do not expect that kinetic terms in the 
effective Lagrangian possess analogous holomorphic properties.
Similarly, holomorphy is apparently of no use 
for correlation functions with non-zero momentum in the full 
theory.}.
Precisely this kind of holomorphy for QCD with light quarks
in the Pauli-Villars fermion mass $ M_{R} $ 
was suggested some time ago
by K\"{u}hn and Zakharov \cite{KZ}
in a somewhat different context. Working in the chiral limit,
these authors have related, using analyticity in $ M_R $, 
the proton matrix element of the topological density  
$ \la p | G \tilde{G} | p \ra $ to the matrix element  
$ \la p | G^2 | p \ra $ which is fixed by the conformal anomaly.
Recently, we have applied \cite{2} a similar idea
to relate the zero momentum two-point function of  
$  G \tilde{G} $ to that of $ G^2 $. The statement of holomorphy in
the fermion mass $ 
m \rightarrow \infty $ was rephrased there as a method to evaluate 
zero momentum correlation
functions of the chiral fermion bilinears
$ \bar{\Psi}_L \Psi_R $ and $ 
  \bar{\Psi}_R \Psi_L $ with the operators $ G^2 , G \tilde{G} $
from the known Ward identities involving the gluon operators only. 
As will be clear below, the value of $ \xi $, the parameter of 
interest, follows from these relations. 
It is our purpose here to show that these results of Ref. \cite{2} 
follow from the effective Lagrangian
constructed in Sect.3. Therefore, it comes as no surprise that
in the effective Lagrangian approach we end up with precisely
the same value of $ \xi $ as that obtained in \cite{2}. (Again,
we recall that the same regularization scheme is implied in the 
approach of \cite{2} and the present paper.)

Performing the inverse Legendre transform of the effective potential
(\ref{22}) 
with respect to $ H $ and $ U $,
we obtain the generating functional $ \tilde{W}_{m}(J, j )
$ of connected Green functions: 
\beq
\label{34}
 \tilde{W}_{m}(J, j ) = - \frac{1}{4} c \Lambda_{QCD}^4 \left( 
\frac{m + j }{ \Lambda_{QCD}} \right)^{8/(3b)} \, e^{-4 J } \; .
\eeq
It is clear from the above analysis that the sources $ j , J $ are 
holomorphic: the differentiation in respect to $ j $ produces 
insertions of the chiral operator $ - \bar{\Psi}_L \Psi_{R} $, 
while the derivative $ \partial / \partial J $ produces insertions
of the operator $ H $. As is seen from Eq.(\ref{34}), the 
differentiation with respect to 
$ j $ can be substituted by the differentiation
with respect to $ m $. In this way we find the correlation function
\beq
\label{36}
i \int dx \lo T \{ H(x) \; m \bar{ \Psi}_L \Psi_R (0) \} 
\ro = - m \frac{ \partial}{ \partial m} \, \frac{ 
 \partial}{ \partial J } \,  \tilde{W}_{m}(J, 0 ) = - \frac{8}{3 b}
\la H \ra \; .
\eeq
Analogously we can find further correlation functions:
\beq
\label{37}
i \int dx \lo T \{ \bar{H}(x) \; m \bar{ \Psi}_R \Psi_L (0) \} 
\ro = - \frac{8}{3 b}
\la \bar{H} \ra \; ,
\eeq
\beq
\label{38}
i \int dx \lo T \{ \bar{H}(x) \; m \bar{ \Psi}_L \Psi_R (0) \} 
\ro = 0 \; ,
\eeq
\beq
\label{39}
i \int dx \lo T \{ H(x) \; m \bar{ \Psi}_R \Psi_L (0) \} 
\ro = 0 \; .
\eeq
The relations proposed in Ref.\cite{2} now follow 
if we take linear combinations of Eqs. (\ref{36}-\ref{39}). 
Taking the sum of Eqs. (\ref{36}) and (\ref{38}) and using
\[  
 H + \bar{H} = - b \alpha_s/(8 \pi) G^2 \; \;  , \; \; 
\la H \ra =  
(3b/2) \la m \bar{ \Psi}_L \Psi_R \ra  \; , \] 
we obtain
\beq
\label{40}
i \int dx \lo T \{ \frac{ \alpha_s}{12 \pi}
G^2(x) \; m \bar{ \Psi}_L \Psi_R (0) \} \ro =  \frac{8}{3 b}
 \la m \bar{ \Psi}_L \Psi_R \ra  \; ,
\eeq
and analogously from Eqs. (\ref{37}),(\ref{39})
\beq
\label{41}
i \int dx \lo T \{ \frac{ \alpha_s}{12 \pi}
G^2(x) \; m \bar{ \Psi}_R \Psi_L (0) \} \ro  =  \frac{8}{3 b}
 \la m \bar{ \Psi}_R \Psi_L \ra  \; .
\eeq
On the other hand, taking the differences for the same pairs 
of Eqs. (\ref{36}-\ref{39}), we obtain
\beq
\label{42}
i \int dx \lo T \{ \frac{ \alpha_s}{8 \pi}
G \tilde{G} (x) \; m \bar{ \Psi}_L \Psi_R (0) \} \ro = 
 i \frac{8}{3 b}
 \la m \bar{ \Psi}_L \Psi_R \ra  \; ,
\eeq
\beq
\label{43}
i \int dx \lo T \{ \frac{ \alpha_s}{8 \pi}
G \tilde{G} (x) \; m \bar{ \Psi}_R \Psi_L (0) \} \ro =  
- i \frac{8}{3 b}
 \la m \bar{ \Psi}_R \Psi_L \ra  \; .
\eeq
Eqs. (\ref{40}-\ref{43}) are precisely the relations suggested 
earlier in Ref.\cite{2} with a different motivation
to find the topological susceptibility in terms of the gluon 
condensate.
Taking the difference of Eqs. (\ref{42}) and (\ref{43}) and using 
Eqs.(\ref{25}), we obtain 
\beq
\label{44}
i \int dx \lo T \{ \frac{ \alpha_s}{8 \pi}
G \tilde{G} (x)  \;  \frac{ \alpha_s}{8 \pi}
G \tilde{G} (0) \} \ro  =  \left( \frac{4}{3 b} \right)^2 
 \la - \frac{b \alpha_s}{8 \pi} G^2  \ra  \; ,
\eeq
which is exactly the result obtained in \cite{2}. (As in \cite{2},
this scheme dependent result implies a particular regularization 
prescription where correlation functions are defined via the path 
integral with perturbative contributions subtracted, see the discussion
above.)   
We have thus closed 
the circle: starting with Eq.(\ref{3}) with an unspecified parameter 
$ \xi $, we have found the value of $ \xi $ using the effective 
Lagrangian, obtained from the anomalous 
Ward identities (\ref{1}-\ref{3}), 
along with the integrating in/out technique, and eventually
fixed the initial Eq.(\ref{3}).   
The introduction of a heavy quark was crucial to find the 
parameter $ \xi $. One should note that the main assumptions
of the present paper and that of Ref.\cite{2} are quite 
different, and yet lead to the same result (\ref{44}). We 
consider this as an evidence in favor of correctness and 
self-consistency of our approach. 

\section{Discussion and comparison with related works}

The purpose of this section is a qualitative discussion of 
the results obtained in the present paper. We will first 
address the counting of vacua following with our methods
and compare its $ N_c $ dependence with the results 
obtained for softly broken SUSY models \cite{MV}
and the behavior found for
the lattice $ Z_p $ gauge models \cite{Cardy}. Another purpose of this 
section is a qualitative comparison of our picture of the 
vacuum structure and $ \theta $ dependence in YM theory with 
a recent work by Witten \cite{Wittheta} who studies the same issues 
in the limit $ N_c \rightarrow \infty $ on the basis of 
the AdS/CFT correspondence \cite{Mald}.
    
As was discussed in \cite{1}, the value of $ \xi $ determines the 
number of different 
nondegenerate $ \theta $ vacuum sectors 
in pure YM theory. When the rational
number $ p/q = 1/(2 \xi ) $ is fixed with $ p $ and $ q $ 
being relatively prime, the number of different 
non-degenerate $ \theta $ 
vacua\footnote{Here we would like to recall that,  for any generic 
value of $ \theta $,
there is only one true physical vacuum which is a lowest energy 
state among 
all $ \theta $ vacua. For a fixed value of $ \theta $, additional
vacua show up in the thermodynamics
limit only indirectly via the value of parameter $ \xi $ in 
the correlation function (\ref{3}).} is 
$ p $. More precisely:
\bea
\label{N}
number~ of~ different ~\theta ~ sectors = p= 
11\cdot N_c~~ for ~ N_c~ =odd  
  \nonumber  \\
  p= 
min. integer [\frac{11N_c}{2}; \frac{11N_c}{4}; 
 \frac{11N_c}{8}]~ for ~~ N_c ~=~even \; ,  
\eea
 where we have used $ \xi = 4/(3 b) $. In particular, it 
follows from Eq.(\ref{N})  
that the number of vacua is 11 for $ N_c = 2 $, and $33$
 for   $ N_c=3  $.
Although these numbers may look strange, they seem to be the only 
ones compatible with the effective Lagrangian and 
integrating in procedure considered in this paper. (They would be 
a wrong answer if one of the assumptions of the present 
approach were incorrect.) 
As was pointed out
in \cite{1}, this counting of vacua disagrees with what could be 
expected starting from SUSY YM theory broken softly 
by a small gluino mass $ m_g \ll \Lambda $ \cite{MV}. The latter theory 
predicts $ N_c $ vacua 
for small $ m_g $. Here is how it comes about (see a discussion by 
Shifman in \cite{SUSY} for more detail).  In the limit of 
small $ m_g $ the 
VEV
of the holomorphic combination $  G^2 + i G \tilde{G} $ is 
proportional to the VEV  $  m_g \la \lambda \lambda \ra $ where the 
gluino condensate $ \la \lambda \lambda \ra $ is to be calculated 
in the supersymmetric limit $ m_{g} = 0 $. The $ \theta $ 
dependence of 
the latter is known \cite{SV}: $ \la \lambda \lambda \ra \sim 
\exp (i 
\theta/ N_c + 2 \pi k /N_c ) \, , k = 0,1, \ldots, N_c -1 $, which 
corresponds to $ N_c $ degenerate vacua. When $ m_g \neq 0 $, 
the vacuum degeneracy is lifted. For $N_c = 3 $ and $ \theta = 0 $, 
we have one state with negative energy $ E = - m_g \Lambda_{SYM}^3
$, and two degenerate states with positive energy $ E = (1/2)    
m_g \Lambda_{SYM}^3 $. The former is the true vacuum state of  
softly broken SUSY gluodynamics, while the latter are metastable 
states with broken $ CP$. The lifetime of the metastable states 
is very large for small $ m_g $, and decreases as $ m_g $ approaches
$ \Lambda_{SYM}$ . When $ \theta $ is varied, the three states 
intertwine,
thus restoring the physical $ 2 \pi $ periodicity in $ \theta $.  
This picture implies the values $ p = N_c $, $ q = 1 $,
different from those suggested by the present approach.

The problem with the above SUSY-motivated scenario is that the genuine
case of pure YM theory corresponds to the limit $ m_g \gg 
\Lambda_{SYM} $ which is not controlled in this 
approach. Thus, although  
naively one could expect that increasing 
of $ m_g  $ to higher values $ m_g \geq \Lambda $ does not change 
the number of vacua of the theory, this 
expectation is unwarranted. It is  conceivable that an 
additional level splitting occurs 
with passing the region  $ m_g \sim \Lambda_{SYM}
$ where the SUSY methods become inapplicable. Our results 
imply that this is indeed what happens, i.e. that 
there exists  a sort of phase 
transition that separates the softly broken SUSY YM theory with 
 $ m_g \ll \Lambda $ from pure non-supersymmetric gluodynamics.
This expectation is consistent with numerous evidences
within the soft SUSY breaking theories indicating that the 
naive decoupling limit  $ m_g \gg 
\Lambda_{SYM} $ produces results incompatible with the known 
infrared features of QCD. In particular, they include the 
wrong $ N_c $ dependence of the effective chiral dynamics \cite{MW}
and the run-away behavior for $ N_f < N_c $ in the presence of 
supersymmetry breaking \cite{Barbon}.  On the other hand, it 
is curious to mention that  
the same value $ p = 11 N_c $ follows within a non-standard 
non-soft SUSY breaking suggested recently \cite{SS} as a toy model
to match the conformal anomaly of non-supersymmetric YM theory
at the effective Lagrangian level.

Our next remark concerns with another feature of the $ N_c $ 
dependence 
in Eq.(\ref{N}). 
Naively, one could expect (and supersymmetric models 
support this expectation)
that small variations of $N_c$ lead to small variations in 
the number of vacuum states.
Our results suggest quite a different picture for 
certain values of $ N_c $: when the number
of colors changes from $N_c=8k-1$ to $N_c=8k$, the number of 
vacuum states
abruptly changes from $11 (8k-1)$ to $11 k$. At the same time,
 for generic values of $ N_c $, the variation of a number of
vacua is quite smooth. Have we ever met with such kind of 
behavior in physics\footnote{
We thank the Referee for asking
this question.}? The answer is yes. When a theory possesses  
two (or more)
relevant integer parameters, the vacuum structure of the theory may 
undergo  very dramatic changes 
with variations of these parameters. As 
an example one could consider 
$Z_p$ lattice gauge models in 4 dimensions in the presence of a $\theta$
term which takes the  
values $ \theta/(2 \pi) = l/q $
where $ l $ and $ q $ are 
integers \cite{Cardy}. In this case the physics is very sensitive
to the numerical value of $q$, namely  
whether it is proportional to $l$ or not. 
The physics changes drastically exactly at the  points when 
$q = l \cdot n  $ where $ n $ is another integer, much like
in our case described above. (The concrete examples when 
$ \theta/(2 \pi) = 1/N $
and $ \theta/(2 \pi) = 2/N $ with $ N $ odd, were considered in detail 
in Ref. \cite{Cardy}.)   
The physical explanation for such a behavior in the 
$Z_p$ model
is based on the existence of a dual description where the relevant 
degrees of freedom  are quite different from the original ones. In 
particular, exactly at these
points one can construct a unique composite field which may condense
according to 
different patterns depending on the value of $ \theta/(2 \pi) $. 

In a sense, the supersymmetric models are similar 
to the $Z_p$ lattice gauge models
with $\theta/(2 \pi) =1/N$ where a very smooth behavior is expected. 
We believe that the
non-supersymmetric models are closer to the case 
when $\theta/(2 \pi)=\frac{l}{q}$
in the $Z_p$ lattice gauge model with $l, q$ being relatively prime. 
We do not know whether the explanation using the dual description
for the $ Z_p $ models
can be extended to the case of gluodynamics, but  
an analogy with the examples
discussed in Ref.\cite{Cardy} suggest that this might be the case.

Finally, we would like to comment on another 
related development.
Very recently, Witten \cite{Wittheta} has shown how the qualitative 
features of the $ \theta $ dependence in non-supersymmetric 
YM theory - such as a multiplicity of vacua $ \sim N_c $,  
existence of domain walls and 
exact vacuum doubling
at some special values of $ \theta $ - can be understood using 
the AdS/CFT duality.  The latter \cite{Mald} 
provides a continuum version of 
the strong coupling limit, with a fixed ultraviolet
cutoff, for YM theory with  
$ N_c \rightarrow \infty $, $g_{YM}^{2} N_c \rightarrow \infty
$. As was shown in \cite{Wittheta},
in this regime the $ \theta $ dependence of the vacuum energy
in YM theory takes the form 
\beq
\label{Wt}
E_{vac} (\theta) = C \, \min_{k} ( \theta + 2 \pi k)^2 + 
O(1/N_c) \; , 
\eeq   
where $ C $ is some constant.
We would like to make two comments on a comparison of
our results  
with the picture advocated by Witten in the large $ N_c $ 
limit. First, we note 
that the structure of Eq.(\ref{Wt}) agrees with our modified 
definition of the path integral including summation over
all branches of a multi-valued (effective) action. Indeed,
Eq.(\ref{Wt}) suggests the correspondence
\beq
\label{Wt2}
C \, \min_{k} \, ( \theta + 2 \pi k)^2  \Leftrightarrow
\lim_{V \rightarrow \infty} \, \left( - \frac{1}{V} \right)
\, \log \left[ \sum_{k} e^{-VC (\theta + 2 \pi k)^2 } \right]
\eeq
using the definition of the vacuum energy through the 
thermodynamic limit of the path integral. With this 
definition which prescribes the way the volume $ V $ appears
in the formula for the vacuum energy, the correspondence
(\ref{Wt2}) appears to be the only possible one. 
On the other hand, the latter expression is 
exactly what arises (in the large $ N_c $ limit)
with our definition of the improved effective potential (\ref{10}).
Therefore, our prescription of summation over
all branches of a multi-valued (effective) action
seems to be consistent with the picture developed by Witten
using an approach based on the AdS/CFT correspondence.
In particular, our picture of bubbles of metastable vacua
bounded by domain walls considered in the context 
of QCD within an effective Lagrangian approach 
in \cite{FHZ} is in qualitative agreement 
with that suggested by Witten \cite{Wittheta}
for the pure YM case. 

Second, one may wonder whether the approach of Ref. \cite{Wittheta}
can provide an alternative way to fix the paramaters $ p,q $ of 
interest. We note that   
Eq.(\ref{Wt}) indicates
a non-analyticity at the 
values $ \theta_c  = \pi \; (mod \, 2 \pi) $ only, where $CP$ 
is broken spontaneously. If the technique based 
on the AdS/CFT duality could be smoothly continued to the weak 
coupling regime of non-supersymmetric YM theory, this would 
result in the values  $ q = 1 , p \sim N_c $. 
However, the possibility of such extrapolation is unclear,
as for small $ \lambda = g_{YM}^{2} N_c $ the 
background geometry develops a singular behavior and the 
supergravity approach breaks down.
%Analogously, the large $ N_c  $ approximation is not sufficient
%for the analysis of stability of different vacua. 
There might well be a phase transition \cite{ft}
when the effective YM coupling $ g_{YM}^2 \, N_{c} $ is reduced.
That such a phase transition should occur in 
the supergravity approach to $ QCD_3 $ was argued 
in \cite{GO}. Other reservations about the use of the supergravity
approach to the non-supersymmetric YM theory in D=4 have been expressed
in \cite{Oog} where no perturbative indication was found for 
decoupling of unwanted massive Kaluza-Klein states of string theory.
On the other hand, there exist some evidences from 
lattice simulations that a critical value of $ \theta $
moves from $ \theta_c = \pi $ in the strong coupling 
regime to $ \theta_c < \pi $ in the weak coupling regime \cite{lattice}. 
In terms of parameters $ p ,q $, such a case corresponds to 
$ q \neq 1 $. Therefore, we conclude that if no phase transition
existed in the supergravity approach, our results would be 
in conflict with the latter which would imply $ p = O(N_c), q = 1 $.
In this case, the assumptions made in the present work would have 
to be reconsidered.
Alternatively, there might be no conflict between the two approaches
if such a phase transition does occur.

\section{Conclusions}

In this paper we suggested using the integrating in/out procedure
to study the properties of a low energy Lagrangian for 
gluodynamics obtained in \cite{1}. We have shown that a particular 
holomorphic structure of this effective Lagrangian 
naturally corresponds to 
holomorphy in the fermion mass in an extended theory with a heavy
fermion, provided the 
standard form of the
fermion mass term is used. This observation supports the 
proposals of Refs. \cite{KZ} and
\cite{2} where the idea of holomorphy in the regulator or heavy
fermion mass was applied, respectively, to the study 
of matrix elements and correlation functions of the topological
density operator $ G \tilde{G} $. We have argued that the integrating
in/out method provides not only a perturbative matching of the 
dimensional scale parameters in different theories, but also has 
a global, ``topological", content. This 
holomorphic and ``topological" matching
of the theories with and without a heavy fermion was shown to fix,
under certain assumptions, the 
number of different non-degenerate $ \theta $ vacua for YM theory.  
The result obtained implies that the number of vacua for softly 
broken SUSY YM gluodynamics changes discontinuously
when the gluino mass becomes large. A check of this conclusion
by different methods and analysis of its possible consequences
would be an interesting problem for a further study.
In particular, we have seen that the analysis of the $ \theta $ 
dependence is relevant for the question of presence or absence
of a phase transition in the supergravity approach to the 
non-supersymmetric YM theory.
In a more general context, it is perhaps worthwhile to point
out that the modified path integral prescription  of summation over
all branches of a multi-valued (effective) action suggested
for the case of pure YM theory in \cite{1} (and previously proposed
in different settings in Refs. \cite{Smilga},\cite{KS})
is consistent with the picture developed by Witten
using an approach based on the AdS/CFT correspondence. 
Analogous modifications of the partition function 
for the case of lattice regularized sigma 
models and Abelian lattice models are found to be the only 
self-consistent method  
\cite{Seb} for the analysis of dualities on the lattice.  
This may provide further indications that such modifications
of the partition function is the correct way to work with 
multi-valued actions.


\clearpage
\begin{thebibliography}{99}
\bibitem{Seib} N. Seiberg, Phys. Lett. {\bf B318} (1993) 469;
Phys. Rev. {\bf D49} (1994) 6857.
\bibitem{SUSY} K. Intriligator and N. Seiberg, hep-th/9509096. \\
M. Peskin, hep-th/9702094. \\
M. Shifman, hep-th/9704114. 
\bibitem{VY} G. Veneziano and S. Yankielowicz, Phys. Lett.
{\bf 113B} (1982) 231. \\
T. Taylor, G. Veneziano and S. Yankielowicz, Nucl. Phys. 
{\bf B218} (1983) 439.
\bibitem{Int} K. Intriligator, R.G. Leigh, and N. Seiberg,
Phys. Rev. {\bf D50} (1994) 1092. \\
K. Intriligator, Phys. Lett. {\bf B336} (1994) 409.
\bibitem{ADS} I. Affleck, M.Dine, and N. Seiberg, Nucl. Phys.
{\bf B241} (1984) 493; Nucl. Phys. {\bf B256} (1985) 557.
\bibitem{IS} L. Intriligator and N. Seiberg, Nucl. Phys. {\bf B431}
(1994) 484.
\bibitem{FP} D. Finnell and P. Pouliot, Nucl. Phys. {\bf B453}
 (1995) 225.
\bibitem{1} I. Halperin and A. Zhitnitsky, Phys. Rev. {\bf D58}
(1998) in press; hep-ph/9711398.
\bibitem{KS} A. Kovner and M. Shifman,  Phys. Rev. {\bf D56} (1997) 
2396; hep-th/9702174.
\bibitem{SV}M.A. Shifman and A.I. Vainshtein,  Nucl. Phys. {\bf B296} 
(1988) 445.
\bibitem{2} I. Halperin and A. Zhitnitsky, Mod. Phys. Lett.
{\bf A13} (1998) 1955; hep-ph/9707286.
\bibitem{Wittheta} E. Witten, hep-th/9807109.
\bibitem{NSVZ} V.A. Novikov, M.A. Shifman, A.I. Vainshtein and 
V.I. Zakharov, Nucl. Phys. {\bf B191} (1981) 301. 
%\bibitem{BPST} A.A. Belavin, A.M. Polyakov, A.S. Schwarz and
%A.S. Tyupkin, Phys. Lett. {\bf 59B} (1975) 85.
%\bibitem{Jac} C. Callan, R. Dashen, and D. Gross, Phys. Lett. 
%{\bf 63B} (1976) 172. \\
 %R. Jackiw and C. Rebbi, Phys. Rev. Lett. {\bf 37}
%(1976) 172.
%\bibitem{tH} G. `t Hooft, Phys. Rev. {\bf D14} (1976) 3432; 
%Phys. Rep. {\bf 142} (1986) 357. 
%\bibitem{Wein} S. Weinberg, Phys. Rev. {\bf D11} (1975) 3583. 
%\bibitem{Wit}  E. Witten, Nucl. Phys. {\bf B156} (1979) 269. \\
% G. Veneziano, Nucl. Phys. {\bf B159} (1979) 213.
%\bibitem{SVZ}M.A. Shifman, A.I. Vainshtein, and 
%V.I. Zakharov, Nucl. Phys. 
%{\bf B166} (1980) 493.. 
%\bibitem{Crewther} R.Crewther, P. Di Vecchia, G. 
%Veneziano, and E. Witten,
 %Phys. Lett. {\bf B88} (1979) 123.
%\bibitem{PQ} R. Peccei and H. Quinn, Phys. Rev. Lett. 
%{\bf 38} (1977) 1440. \\
%S. Weinberg, Phys. Rev. Lett. {\bf 40} (1978) 223. \\
%F. Wilczek, Phys. Rev. Lett. {\bf 40} (1978) 279. 
%\bibitem{Kim} J.E. Kim, Phys. Rev. Lett. {\bf 43} (1979) 103.  
%\bibitem{ARZ}A.R. Zhitnitsky, Yad.Fiz. {\bf 31} (1980) 497 
%(Sov. J. Nucl. Phys.
%{\bf 31} (1980) 260).  
%\bibitem{DFS}M. Dine, W. Fischler, and M. Srednicki, Phys. 
%Lett. {\bf B104}
%(1981) 199. 
%\bibitem{axion} J.E. Kim, Phys. Rep. {\bf 150} (1987) 1. \\
%H.Y. Cheng, Phys. Rep. {\bf 158} (1988) 1.
%\bibitem{Smilga}A.V. Smilga, Phys. Rev. {\bf D49} (1994) 6836.
%\bibitem{KSS} A. Kovner, M. Shifman, and A. Smilga, hep-th/9706089.
%\bibitem{Cr} R.J. Crewther, Phys. Lett. {\bf 93B} (1980) 75.
\bibitem{Wit2} E. Witten, Ann. Phys. {\bf 128} (1980) 363. \\
 P. Di Vecchia and G. Veneziano, Nucl. Phys. {\bf B171} 
(1980) 253.
%\bibitem{Wit}  E. Witten, Nucl. Phys. {\bf B156} (1979) 269. \\
 %G. Veneziano, Nucl. Phys. {\bf B159} (1979) 213.
\bibitem{QCD} I. Halperin and A. Zhitnitsky, hep-ph/9803301.
%\bibitem{VV} P. Di Vecchia and G. Veneziano, Nucl. Phys. {\bf B171} 
%(1980) 253.
%\bibitem{Dash} R.F. Dashen, Phys. Rev. {\bf D3} (1971) 1879.

\bibitem{KZ}  J.H. K\"{u}hn and V.I. Zakharov,
Phys. Lett. {\bf B252} (1990) 615.
\bibitem{MV} A. Masiero and G. Veneziano, Nucl. Phys. 
{\bf B249} (1985) 593.
\bibitem{Cardy} J. Cardy and E. Rabinovici, Nucl. Phys. {\bf B205} 
(1982)1. \\ 
J. Cardy,  Nucl. Phys. {\bf B205} 
(1982)17.
\bibitem{Mald} J. Maldacena, hep-th/9711200. \\
S. Gubser, I.R. Klebanov and A.M. Polyakov, hep-th/9802109. \\
E. Witten, hep-th/9802150, hep-th/9803131.
\bibitem{MW} S. Martin and J. Wells, hep-th/9801157.
\bibitem{Barbon} J.L.F. Barbon and A. Pasquinucci, hep-th/9804029.
\bibitem{SS}  F. Sannino and J. Schechter, hep-th/9708113.
\bibitem{ft} D. Gross and H. Ooguri, hep-th/9805129. 
\bibitem{GO} J. Greensite and P. Olesen, hep-th/9806235.
\bibitem{Oog} H. Ooguri, H. Robins and J. Tannenhauser, hep-th/9806171.
\bibitem{lattice} G. Schierholz, Nucl. Phys. Proc. Suppl. {\bf 42}
(1995) 270; hep-lat/9412083. \\
A.S. Hassan, M. Imachi, N. Tsuzuki and H. Yoneyama, Prog. Theor. Phys.
{\bf 95} (1996) 175. \\ 
H. Yoneyama, talk at Lattice-98.
\bibitem{FHZ} T. Fugleberg, I. Halperin and A. Zhitnitsky,
to appear.
\bibitem{Smilga}A.V. Smilga, Phys. Rev. {\bf D49} (1994) 6836.
\bibitem{Seb} S. Jaimungal, hep-th/9805211, hep-th/9808018.
%\bibitem{Sch} J. Schechter, Phys. Rev. {\bf  D21} (1980) 3393.
%\bibitem{MS} A.A. Migdal and M.A. Shifman, Phys. Lett. {\bf 114B}
%(1982) 445. See also M.A. Shifman, Phys. Rep. {\bf 209} (1991) 341. 
%\bibitem{Verg} S.N. Vergeles, Nucl. Phys. {\bf B152} (1979) 330.
%\bibitem{Pol} A.M. Polyakov, Nucl. Phys. {\bf B120} (1977) 429.
%\bibitem{Kon} M. Di Pierro and K. Konishi, Phys. Lett. {\bf B388}
%(1996) 90 (hep-th/9605178). \\
%N. Evans, S. Hsu, and M. Schwetz, Nucl. Phys. {\bf B484} (1997)
%124 (hep-th/9608135). \\
%K. Konishi, Phys. Lett. {\bf B392} (1997) 101 (hep-th/9609021).
%\bibitem{SW}  N. Seiberg and E. Witten, Nucl. Phys. {\bf B426}
%(1994) 19; {\bf B431} (1994) 484.
%\bibitem{CS} J. Cornwall and A. Soni, Phys. Rev. {\bf D29} (1984) 1424.
%\bibitem{Creutz} M. Creutz, Phys. Rev. {\bf D52} (1995) 2951. \\
%N. Evans, S. Hsu, A. Nyffeler, and M. Schwetz, Nucl. Phys. {\bf B494}
%(1997) 200.
%\bibitem{tHooft} G. `t Hooft, Nucl. Phys. {\bf B190} (1981) 455.
%\bibitem{sigma}  V.A. Novikov, M.A. Shifman, A.I. Vainshtein and 
%V.I. Zakharov, Phys. Rep. {\bf 116} (1984) 103.
%\bibitem{DE} D.I. Dyakonov and M.I. Eides, Sov. Phys. JETP
%xdvu{\bf 54} (1981) 232.
%\bibitem{VZ} A.I. Vainshtein and 
%V.I. Zakharov, Sov. Phys. JETP {\bf 68} (1989) 701; Nucl. Phys.
%{\bf B324} (1989) 495.
%\bibitem{witten1}  E. Witten, hep-th/9706109
% \bibitem{shifman1}  M.A. Shifman, hep-th/970414
%\bibitem{hanany} A. Brandhuber, J. Sonnenschein, S. Theisen,
%and S. Yankielowicz, hep-th/9704044.\\
% A. Hanany, M.Strassler, and A. Zaffaroni, hep-th/9707244. \\
%N. Evans and M. Schwetz, hep-th/9708122.\\
%J.L.F.Barbon and A. Pasquinucci, hep-th/9711030. \\
%N. Evans hep-th/9801159.\\
 
\end{thebibliography}
\end{document}